\documentclass{article}
\usepackage{arxiv}

\usepackage[numbers]{natbib}
\usepackage{subcaption}
\usepackage{graphicx} 
\usepackage{dirtree}
\usepackage{array}

\newcolumntype{M}[1]{>{\centering\arraybackslash}m{#1}}

\title{X-ray and Visible Spectra Circular Motion Images Dataset}

\author{
  Mikhail O.~Chekanov \\
  Institute for Information Transmission Problems\\ 
  (Kharkevich Institute)\\
  Moscow Institute of Physics and Technology\\
  (National Research University)\\
  \texttt{chekanov@visillect.com} \\
  \And
  Oleg S.~Shipitko \\
  Institute for Information Transmission Problems\\
  (Kharkevich Institute)\\
  \texttt{shipitko@visillect.com} \\
}

\begin{document}

\maketitle

\begin{abstract}
We present the collections of images of the same rotating plastic object made in X-ray and visible spectra. Both parts of the dataset contain $400$ images. The images are maid every $0.5$ degrees of the object axial rotation. The collection of images is designed for evaluation of the performance of circular motion estimation algorithms as well as for the study of X-ray nature influence on the image analysis algorithms such as keypoints detection and description. The dataset is available at \texttt{https://github.com/Visillect/xvcm-dataset}. 

\end{abstract}

\section{Introduction}

Computed tomography is widely used in various fields: medicine~\citep{kesminiene2018cancer}, precise measurements~\citep{buratti2018applications}, agriculture~\citep{mairhofer2017x}.
In tomography, the mutual trajectories of the sample, detector, and probe radiation source are usually considered known, since they are determined by the targeted movement of the setup components.
Most of the computed tomography reconstruction methods rely on the geometric accuracy of the instrument and the reliably known trajectories of all its parts~\citep{feldkamp1984practical, katsevich2004improved}
However, the realized trajectory differs from the desired one for various reasons (mechanical backlash, an error in measuring the angle of rotation of an object, thermal deformations, the slope of the sample relative to the axis of rotation), which negatively affects the quality of the reconstruction.
Thus, geometric errors are one of the main sources of reconstruction errors~\citep{ferrucci2015towards}.
To compensate for these deviations different calibration approaches are used. These approaches can be divided into two classes. The first class of geometric calibration methods is based on observations of a specific object with a known geometry~\citep{dewulf2013uncertainty, hermanek2017optimized, weiss2012geometric} or is based on reference measuring instruments~\citep{welkenhuyzen2014investigation, bircher2018geometry}. Geometric calibration is a laborious and expensive process, requiring the involvement of specialists with the appropriate qualifications.
Moreover, geometrical errors inherent even in a calibrated system can still have a negative effect on the quality of reconstruction, and the quality of calibration decreases with time. The second class of methods called online calibration refines trajectory directly during measurements by analyzing obtained projections~\citep{yang2017direct, xu2017simultaneous, zhang2014iterative, muders2014stable, chung2018tomosynthesis}. The online calibration methods do not require additional experiments and allow to compensate for geometric errors. Since circular motion estimation is of practical interest in tomography \citep{ferrucci2015towards} we focus on this particular type of motion. While there are existing datasets for camera motion estimation with images in visible spectrum \citep{hodan2017t, ovg2009multi, seitz2006comparison}, there is no analogous dataset containing X-ray images. Moreover, the application of classical computer vision techniques to digital X-ray images requires an analysis of the influence introduced by the different nature of such images to the quality of algorithms. Again, еhere is no known dataset allowing to assess the influence of the translucent world model applicable for X-ray images on the algorithms developed for the classical opaque world model. 

We present a dataset with digital X-ray images of a plastic object. The purpose of the presented dataset is twofold. It allows to measure the performance of circular motion estimation algorithms on X-ray images as well as to study the difference in computer vision algorithms performance (e.g. keypoints detection and matching) while applied to visible and X-ray data.

\section{Data description}

The presented dataset consists of two parts: (i) images made in visible and (ii) X-ray spectra. The structure of the dataset is presented below:

\begin{minipage}{3in}
\dirtree{%
.1 xvcm\_dataset.
.2 xray.
.3 raw.
.4 dark.
.4 data.
.4 empty.
.3 preprocessed.
.4 downscaled.
.3 calib.
.2 visible.
.3 raw.
.3 preprocessed.
.4 cropped.
.3 calib.
}
\end{minipage}

\begin{figure}[ht]
\centering
\begin{subfigure}[b]{0.7\textwidth}
   \includegraphics[width=1\linewidth]{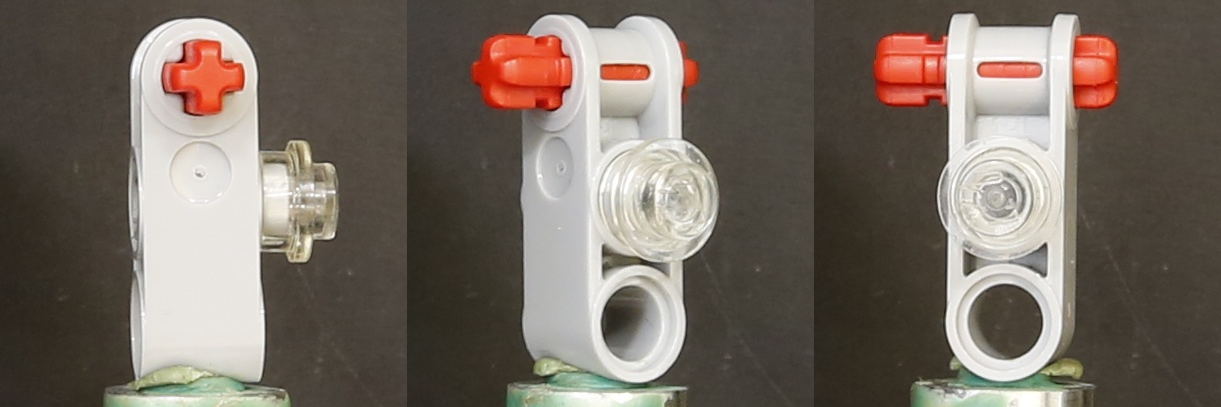}
   \caption{visible spectrum}
   \label{fig:Ng1} 
\end{subfigure}

\begin{subfigure}[b]{0.7\textwidth}
   \includegraphics[width=1\linewidth]{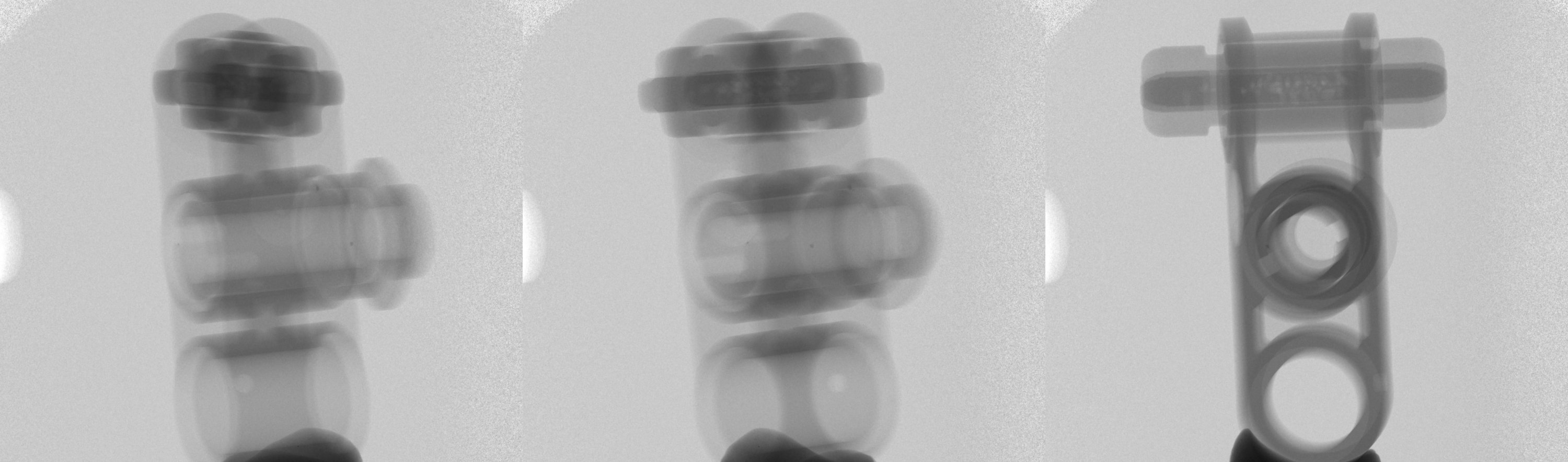}
   \caption{X-ray spectrum}
   \label{fig:Ng2}
\end{subfigure}
\caption{Examples of images presented in dataset for visible (a) and X-ray (b) spectrum.}
\label{fig:data}
\end{figure}

\begin{figure}[ht]
\centering
   \includegraphics[width=0.7\linewidth]{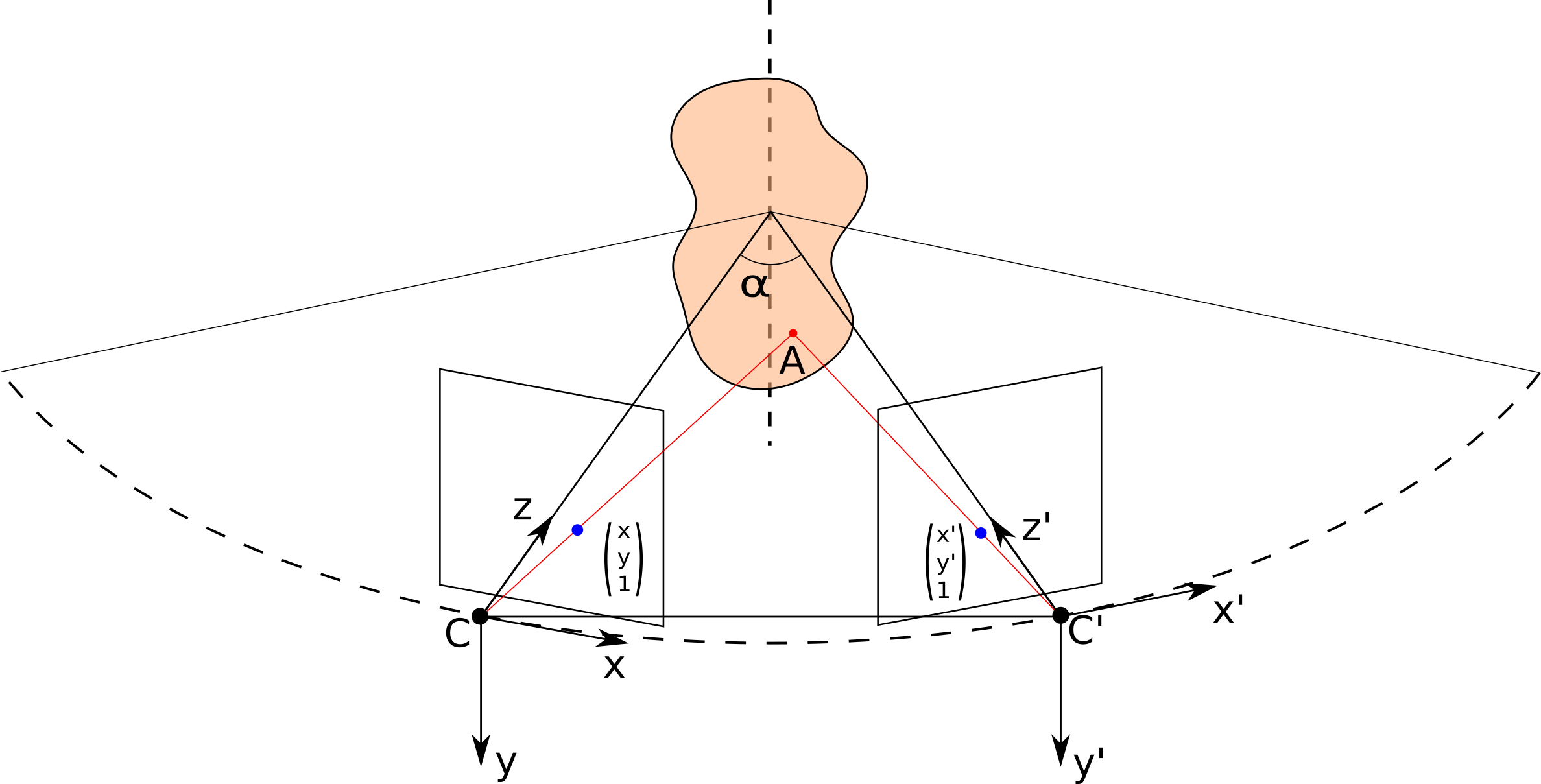}
   \caption{Circular camera motion. $C$ and $C'$ stand for camera centers in different moments of time, $\alpha$ is a rotation angle, $A$ is a point belonging to the observed object.}
   \label{fig:circular_motion} 
\end{figure}

Figure~\ref{fig:data} demonstrates samples images while Table~\ref{tab:params} presents the dataset parameters. The X-ray part of the dataset was collected at the X-ray microtomograph developed and operating at the Federal Research Center for Crystallography and Photonics of Russian Academy of Sciences~\citep{buzmakov2018laboratory, buzmakov2019laboratory} with the following parameters:
\begin{itemize}
    \item exposure time -- 5 seconds;
	\item radiation energy -- 17 keV;
	\item monochromator -- pyrographite;
	\item MoKa line. 
\end{itemize}
While recording data the camera was stationary. The optical axis of the camera was perpendicular to the axis of object rotation. Such system setup is equivalent to the case of the circular camera motion around a stationary object (see Fig.~\ref{fig:circular_motion}).

\subsection{Data preprocessing}
To compensate for X-ray detector noise the following prepossessing algorithm was applied:
\begin{equation}
    \textrm{preprocessed\_image}_{i,j} = \frac{\textrm{data}_{i,j}-mean(\textrm{dark}_{i,j})}{mean(\textrm{empty}_{i,j})-mean(\textrm{dark}_{i,j})}
\end{equation}
where $i, j$ -- pixel coordinates, $mean()$ -- is the mean value of pixels in the given coordinates over the whole subfolder, three subfolders contain the following images:
\begin{itemize}
    \item \texttt{dark} -- images taken in absence of X-ray radiation;
    \item \texttt{empty} -- images taken in absence of object;
    \item \texttt{data} -- images with rotating object.
\end{itemize}    
Prerocessed images are stored in \texttt{xray\textbackslash preprocessed}.

Due to the Canon EOS 5D Mark III camera limitations object of interest occupies only a small part of visible images. Folder \texttt{visible\textbackslash preprocessed\textbackslash cropped} contains cropped images of the size 407 x 407 pixels. To make visible and X-ray images comparable the latest were downscaled to the comparable resolution (407 x 360 pixels) in a aspect ratio preserving way (folder \texttt{visible\textbackslash preprocessed\textbackslash cropped}).

\begin{table}
    \caption{Parameters of visible and X-ray dataset parts}
    \label{tab:params}
    \begin{center}
    \begin{tabular}[h]{|c|M{3.5cm}|c| }
         \hline
         \hline
         \textbf{Parameter} & \textbf{Visible} & \textbf{X-ray} \\ \hline
         Number of images & 400 & 400 \\ \hline
         Rotation angle step & $0.5^\circ$ & $0.5^\circ$ \\ \hline
         Resolution & 5760 x 3840 pixels (object occupies an area of approximately 400 x 400 pixels) & 3000 x 2650 pixels \\ \hline
         Camera & Canon EOS 5D Mark III & Ximea xiRay11 \\
         \hline
         \hline
    \end{tabular}
    \end{center}
\end{table}

\subsection{Calibration}

Both parts of the dataset are accompanied by camera calibration parameters. In case of visible data, calibration was performed with the OpenCV toolbox~\citep{bradski2000opencv} by detecting chessboard pattern. Calibration parameters as well as images used for calibration are stored in \texttt{visible\textbackslash calib} folder. In the case of X-ray data calibration was obtained by the known dimensions of the object were used to estimate camera calibration parameters. The calibration is stored at \texttt{xray\textbackslash calib}.


\section{Conclusion}

We presented the dataset containing both X-ray and visible images of the same rotating object. The presented dataset can be used while developing camera circular motion estimation techniques for online calibration methods. The second purpose of the presented dataset is the evaluation of the computer vision techniques applied to the X-ray data.

\bibliographystyle{unsrt}
\bibliography{bibliography}
\end{document}